# Raman lasing and Fano lineshapes in a packaged fiber-coupled whispering-gallery-mode microresonator

Guangming Zhao[1], Şahin Kaya Özdemir[1], Tao Wang[2], Linhua Xu[1], Gui-lu Long[2], and Lan Yang[1]

[1] *Department of Electrical and Systems Engineering, Washington University, St. Louis, MO 63130;*

2 *State Key Laboratory of Low-Dimensional Quantum Physics and Department of Physics, Tsinghua University,*

*Beijing 100084, People's Republic of China*

We report Raman lasing and the optical analog of electromagnetically-induced-transparency (EIT) in a whispering-gallery-mode (WGM) microtoroid resonator embedded in a low refractive index polymer matrix together with a tapered fiber coupler. The Raman laser supports both single and multimode operation with low power thresholds. Observations of Fano and EIT-like pheonomena in a packaged microresonator will enable high resolution sensors and can be used in networks where slow light process is needed. These results open the way for portable, robust, and stable WGM microlasers and laser-based sensors for applications in various environments.

Whispering gallery mode (WGM) resonators have received increasing interest in many fields of contemporary photonics, such as optical sensing [1-3], nonlinear [4-6] and quantum optics [7,8], lasing [9-13], and optomechanics [14,15] owing to their high quality factors and highly confined fields (i.e., micro-scale mode volume). Traditionally prisms, angle-polished fibers, and tapered fibers have been used for coupling light in and out of WGM microresonators. With advances in fabrication technologies, monolithic fabrication of resonators and their coupling waveguides have also been demonstrated [16]. More recently, there have been reports of coupling free-space light into the resonator by breaking the circular symmetry of WGM resonators via intentionally-induced deformations [17,18] or scatterers [19]. Among these various coupling schemes, tapered fibers have been demonstrated to be ideal couplers because of their high coupling efficiency and their ability to achieve critical coupling where transmission drops to zero [20]. Despite their ideality, there are still issues to be solved to utilize tapered fiber coupling in practical and in-field applications for resonators, because (i) achieving and maintaining a good coupling require the use of expensive nanopositioning systems for each of the resonator that will be used, (ii) fiber tapers are fragile and thus need careful handling when the system is moved out of laboratory, (iii) maintaining long term stability of coupling is difficult (i.e., air flow, mechanical perturbations, etc. will alter the coupling conditions) and requires additional electronics and equipments, and (iv) in an uncontrolled environment out of the laboratory the contaminants may fall on the coupler inducing additional losses or changing the coupling condition. In some environments keeping the resonators clean and free from contaminants is also difficult. Packaging the resonators with their tapered-fiber couplers using a low refractive index polymer has been proposed and demonstrated as a possible remedy for the problems listed above [21].

Spot packaging, where the tapered fiber and the resonator are attached to each other at a spot either by thermally-fusing them together [22] or by gluing the fiber to the resonator using a low-index optical glue [21], has been demonstrated for microspheres [23]. Although this provided a stable coupling, it did not protect the coupler and the resonator from contaminants. In parallel to this, full packaging, where the resonator and tapered fiber are embedded together in a low-index polymer matrix, has been demonstrated for microsphere [21], microtoroid [24]. There are many applications that can benefit from such a packaged resonator system, such as portable WGM microlasers, temperature and humidity sensors, and filters for communication networks.

In this Letter, we demonstrate for the first time Raman lasing and the optical analog of EIT and Fano resonances in a silica microtoroid resonator which is embedded in a low-index polymer matrix together with its coupling tapered-fiber. In these fully packaged systems, we have observed quality factors as high as $2 \times 10^7$ and achieved different coupling conditions, including critical coupling.

We fabricated the silica microtoroid resonators (180 μm major diameter) used in this study according to previously published procedures [25]. During fabrication, we also fabricated reflowed side wall next to each microtoroid [26]. The side walls provide a support for the fiber taper waveguides to rest on so that the fibers do not experience a significant change in their position during the injection and curing of the low-index polymer. We fabricated the tapered fiber using the heat-and-pull method. The light from a tunable laser was coupled into and out of the resonator via this fiber taper. The wavelength of the laser was scanned linearly to obtain the transmission spectra and characterize the resonances. We characterized the resonances using two different tunable lasers one of which was in the 780 nm band and the other was in the 980 nm band. After the resonances were characterized and the proper coupling condition was obtained, we started the packaging process which is outlined schematically in Fig. 1. In our previous work on microtoroid packaging, we used a UV curable low-index polymer (My133MC). Despite the successful implementation, we experienced two major problems. First, the curing process was so fast that the induced tension and stress made it hard to maintain the initially set coupling condition and did not allow time for tuning the coupling condition. Thus, the efficiency of packaging was low. Second, the curing process required that the polymer be isolated from oxygen, so, the packaging had to be done either in a nitrogen box or between two glass slides. After the curing process, it was difficult to separate the packaged chip from the glass slides. In this work, instead of the UV curable polymer, we used a moisture curable polymer (low-loss polymer with a refractive index of 1.33) which needed a longer time for complete curing but enabled us to monitor and continuously tune the fiber-resonator coupling during the process. In this way, we could achieve critical coupling between the fiber and the resonator. In Fig. 1e, we show a typical transmission spectrum obtained for a packaged resonator at critical coupling. The resonance had a quality factor of $10^6$

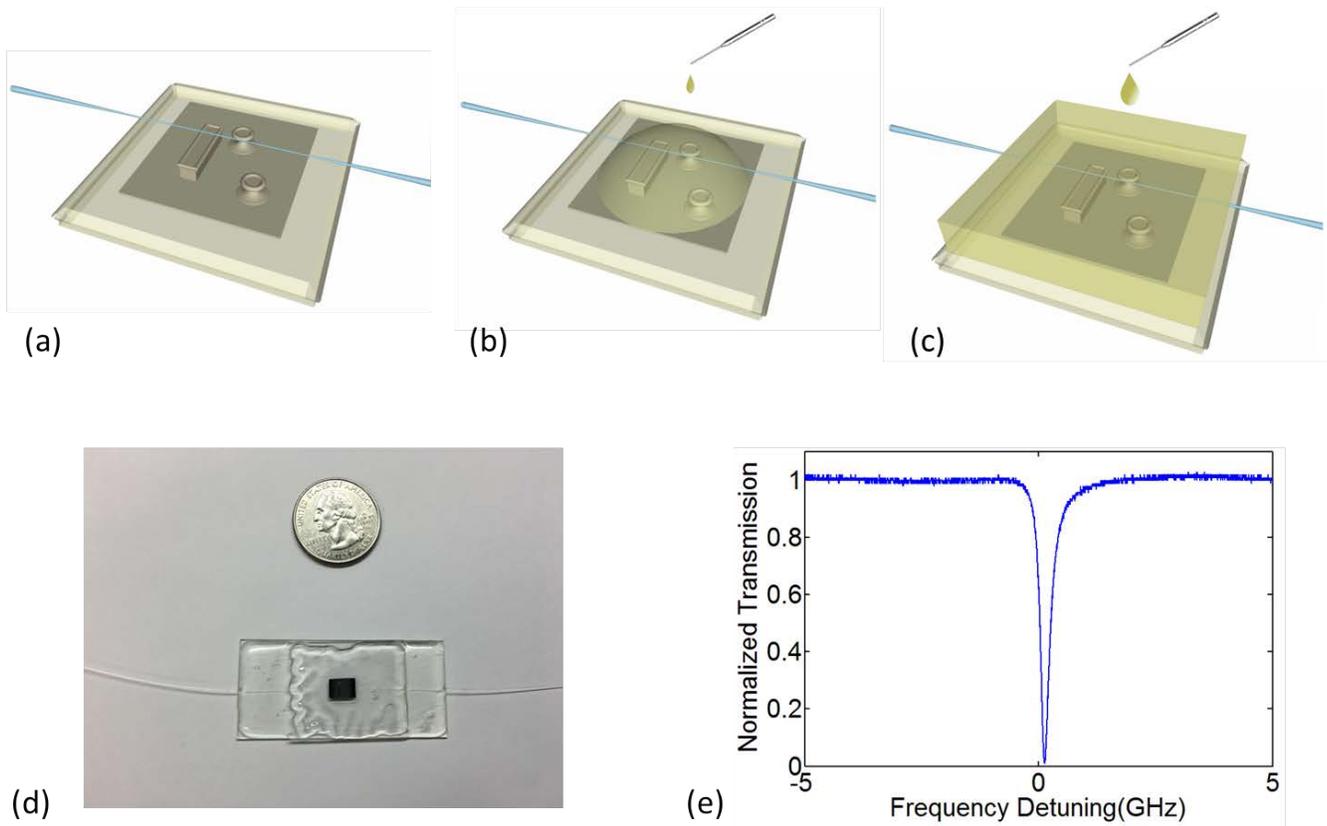

Fig. 1. Packaging a fiber-coupled microtoroid resonator with a low index polymer. (a) The desired coupling condition was first established in air with the help of the reflowed side wall. (b) Polymer was dropped onto the chip in amounts of little droplets, while the transmission was continuously monitored. Any change in the coupling condition, was thus detected, and coupling was accordingly optimized. The process continued until stable transmission and coupling was maintained during the curing process. (c) More polymer was added to complete the packaging process and cover the tapered portions of the fiber completely. (d) An image of the completely cured packaged fiber-coupled resonator. (e) A typical transmission spectrum obtained from one of the packages, depicting critical coupling and a quality factor of $10^6$.

Raman gain in silica microresonators has been used for WGM Raman lasing as well as loss compensation to improve the quality factor. WGM Raman microlasers are very promising for extending the wavelength of existing laser sources. Their use so far has been limited due to the portability issues outlined above. Obtaining Raman gain and Raman laser in a packaged fiber-coupled silica microtoroid will extend the use of these microlasers beyond the laboratory and will certainly benefit many applications. Using the packaged silica microresonator whose transmission spectrum in the pump band of 780 nm is shown in Fig. 2a, we obtained Raman lasing in the 820 nm band. This resonance had a quality factor of $2 \times 10^7$. By adjusting the pump power, we could obtain both single- and multi-mode operations as shown in Fig. 2b, respectively. The lasing threshold is 430µW.

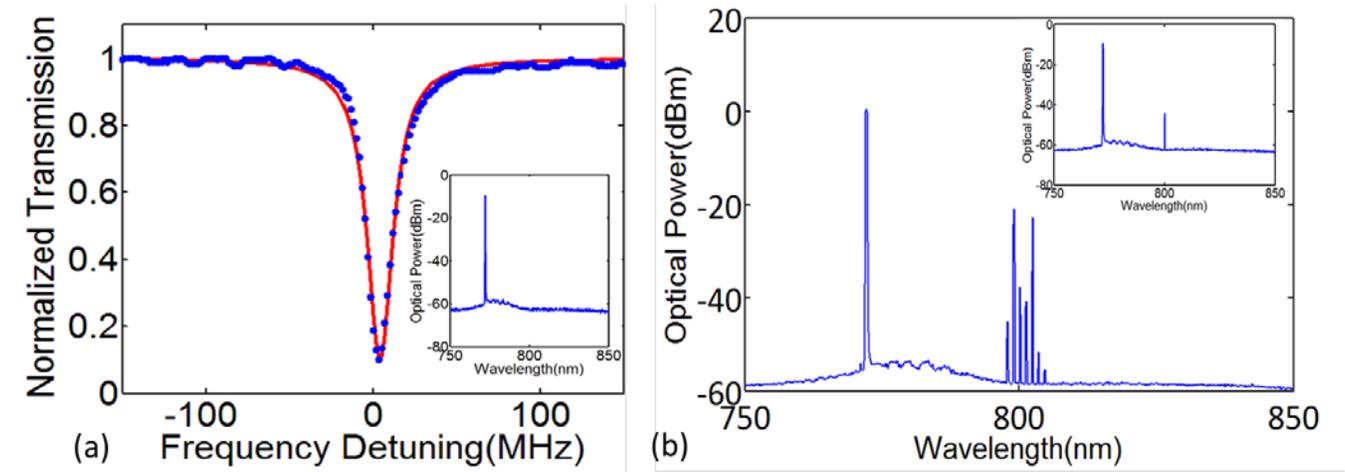

Fig. 2. Raman lasing in a packaged WGM silica microresonator. (a) Transmission spectrum of the packaged microresonator depicting the resonance mode with Q~$2 \times 10^7$ in the pump wavelength band of 780 nm. Inset shows Optical spectrum of pump laser below threshold (b) Optical spectra of single mode Raman lasing (inset) and multimode lasing. The pump is located at 770 nm and the Raman emission is located around 800 nm band.

*Electromagnetically induced transparency (EIT)* and Fano lineshapes have various applications, such as ultraslow light propagation and light storage and as highly sensitive sensors. All-optical analogs of EIT and Fano lineshapes have been demonstrated in various WGM optical microresonators by coupling two optical modes with the same resonance frequency but significantly different quality factors [27-29]. In the packaged resonator we used for Raman lasing, we identified two closely located WGMs, one with low Q~$10^5$ and the other with high Q~$10^6$ (see Fig. 3a), and used them to demonstrate, for the first time, EIT in a packaged WGM resonator, which can maintain a long-term stable coupling condition. The high-Q mode was a low order mode, whereas the low-Q mode was a high order mode, which has a larger field portion in the polymer (Fig. 3b). When the temperature of the packaged system was increased using a thermo-electric cooler, we observed that both of the modes experienced blue shift. However, the low-Q mode shifted faster than the high-Q mode, due to the large negative thermo-optic coefficient of the polymer, in which the low-Q mode had a substantial field portion. The shift in high-Q mode was slower because it was dominantly located in the silica (positive thermo-optic coefficient), with only a small field portion inside the polymer (negative thermo-optic coefficient). This difference in the thermal response of these two modes helped us to vary the detuning between the two modes. The spectrally overlapped modes are then coupled to each other via the fiber taper coupler, resulting in the observation of Fano and EIT-like lineshapes, depending on the frequency detuning between the modes.

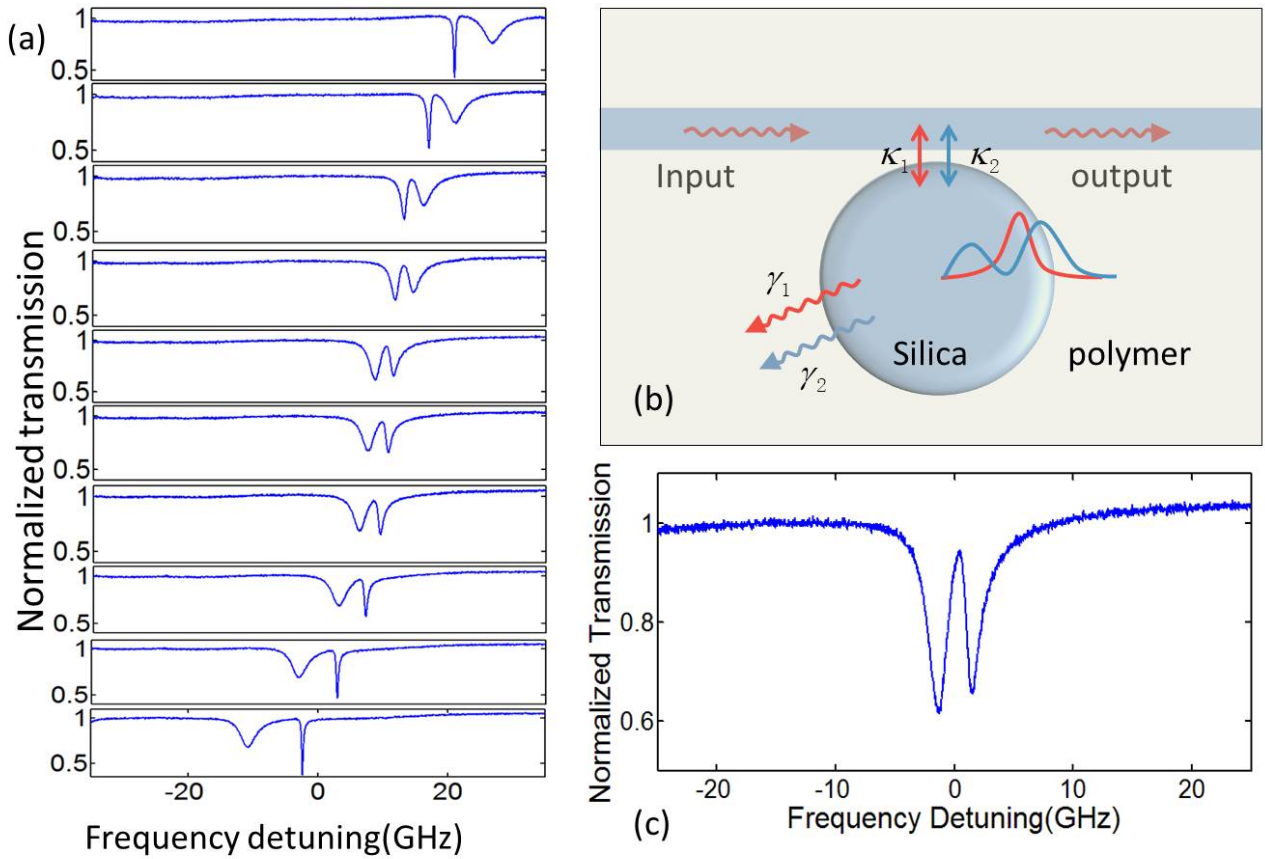

Fig. 3. Fano and EIT-like lineshapes in a packaged microtoroid resonator. (a) Evolution of the transmission spectra obtained in experiments with changes in temperatures. As the temperature increases, the high-Q mode which is initially 8.5 GHz blue-detuned from the low-Q mode becomes 5.5 GHZ red-detuned. Temperature increases from the top spectrum to the bottom spectrum. (b) Schematic illustration of the distribution of the fields of the low- and high-modes in the packaged resonator. The modes with decay rates of $\gamma_1$ and $\gamma_2$ are coupled to the fiber-taper waveguide with different coupling coefficients denoted by $\kappa_1$ and $\kappa_2$. Note that the low-Q mode has a higher field distribution in the polymer. (c) Typical EIT-like spectrum obtained in the experiments.

In conclusion, we have demonstrated an effective way to embed microresonators in a low-index polymer together with their coupling tapered-fibers, and achieved high quality factors and critical coupling. These led to the observation of single and multimode Raman lasing from a packaged silica microtoroid resonator. By making use of the different thermal responses of two resonant modes due to their difference in the field distributions in a packaged microtoroid, we demonstrated Fano and EIT-like lineshapes. We believe that our work will provide stability, robustness, and portability to WGM microresonators and will pave the way to move fiber-coupled WGM microresonators from the laboratory to the field for real-life applications, such as sensing and slow light for communication.